\documentstyle[twocolumn,aps,prl]{revtex} 
 
\draft 
\begin{document} 
 
%%%%%%%%%%%%%%%%%%%%%%%%%%%%%%%%%%%%%%%%%%%%%%%%%%%%%%%%%%%%%%%%%%%%%%%%%%%%%%% 

\title{$\mu^+$-Knight Shift Measurements 
in U$_{0.965}$Th$_{0.035}$Be$_{13}$ Single Crystals}

\author{J.E.~Sonier$^1$, R.H.~Heffner$^1$, D.E.~MacLaughlin$^2$,
G.J.~Nieuwenhuys$^3$, O.~Bernal$^4$, 
R.~Movshovich$^1$, P.G.~Pagliuso$^1$, J.~Cooley$^1$, J.L.~Smith$^1$ 
and J.D.~Thompson$^1$}
\address{$^1$Los Alamos National Laboratory,
Los Alamos, New Mexico 87545}
\address{$^2$Department of Physics, University of California, Riverside, California 92521}
\address{$^3$Kamerlingh Onnes Laboratory, Leiden University, P. O. Box 9506, 
2300 RA Leiden, The Netherlands}
\address{$^4$Physics Department, California State University, 
Los Angeles, California 90032}
\date{February, 2000} 
\date{ \rule{2.5in}{0pt} } 
 
\maketitle 
%%%%%%%%%%%%%%%%%%%%%%%%%%%%%%%%%%%%%%%%%%%%%%%%%%%%%%%%%%%%%%%%%%%%%%%%%%%%%%% 
\begin{abstract} \noindent 

Muon spin rotation ($\mu$SR) measurements of the temperature dependence
of the $\mu^+$-Knight shift in single crystals of U$_{0.965}$Th$_{0.035}$Be$_{13}$
have been used to study the static spin susceptibility $\chi_s$ 
below the transition temperatures $T_{c1}$ and $T_{c2}$. While an abrupt
reduction of $\chi_s$ with decreasing temperature is observed below $T_{c1}$,
$\chi_s$ does not change below $T_{c2}$ and remains at a value below the 
normal-state susceptibility $\chi_n$. In the normal state we find 
an anomalous anisotropic temperature dependence of the transferred 
hyperfine coupling between the $\mu^+$-spin and the U $5f$-electrons.

\end{abstract} 
\pacs{74.70.Tx, 75.30.Mb, 76.75.+i} 
%%%%%%%%%%%%%%%%%%%%%%%%%%%%%%%%%%%%%%%%%%%%%%%%%%%%%%%%%%%%%%%%%%%%%%%%%%%%%%% 
%%\newpage 

An intriguing feature of the heavy-fermion compound U$_{1-x}$Th$_x$Be$_{13}$ 
is that for $0.019 \! \lesssim \! x \! \lesssim \! 0.045$
a second phase transition $T_{c2}(x)$ appears at a temperature below the
superconducting (SC) transition $T_{c1}(x)$ \cite{Ott:85}. 
The nature of the lower transition $T_{c2}$ 
is still a matter of considerable debate. Initially $T_{c2}$
was identified as a second distinct SC transition
from measurements of a specfic heat peak \cite{Ott:85},
the pressure dependence of $T_c(x)$ \cite{Lambert:86}
and the increased slope of $H_{c1}$ vs. $T$ \cite{Rauch:87}. 
The observation of a $T^3$ dependence of the $^9$Be NMR spin-lattice 
relaxation rate below $T_{c2}$ suggested a 
SC state characterized by line nodes in the energy gap 
\cite{MacLaughlin:84}. Later zero-field $\mu$SR
measurements \cite{Heffner:90} clearly revealed the onset of small 
moment magnetism ($\approx \! 10^{-3}$~$\mu_{B}$/U) 
below $T_{c2}$. The appearance of a small
internal field could arise from a SC state 
that breaks time-reversal symmetry \cite{Sigrist:89}. 
On the other hand, it could originate from a spin-density
wave instability \cite{Machida:87} or the 
formation of long-range AFM correlations \cite{Batlogg:85,Kromer:98}  
within a single SC phase. However, these latter 
interpretations fail to explain \cite{Heffner:96} the large specific 
heat jump at $T_{c2}$. 

Early $\mu$SR measurements on polycrystalline 
samples of U$_{0.967}$Th$_{0.033}$Be$_{13}$ showed a constant or 
perhaps weakly increasing $\mu^+$-Knight shift upon cooling
below $T_{c1}$ \cite{Heffner:86}. In the SC state the
temperature dependence of the Knight shift $K$ reflects
the change in the static spin susceptibility $\chi_s$ 
due to the formation of Cooper pairs. 
For the case of orbital $s$-wave 
($L \! = \! 0$) spin singlet ($S \! = \! 0$) pairing, Yosida \cite{Yosida:58}
calculated from the BCS theory that $\chi_s(T)$ vanishes as
$T \! \rightarrow \! 0$~K. Modifications to this 
temperature dependence are expected for spin-orbit scattering by impurities
and unconventional pairing states.
  
In this Letter we report on the temperature dependence of the $\mu^+$-Knight
shift in {\em single crystals} of U$_{0.965}$Th$_{0.035}$Be$_{13}$. 
These measurements differ from earlier studies on polycrystalline
samples in that there are {\em two} magnetically 
inequivalent $\mu^+$-sites which facilitate a determination of $\chi_s$ 
in the SC state. We find that upon cooling through $T_{c1}$, $\chi_s$
rapidly decreases, but remains independent of temperature
below $T_{c2}$. Our study also reveals a temperature dependence in the normal state
of the transferred hyperfine coupling at one of the two $\mu^+$-sites, 
roughly coinciding with features observed in resistivity 
and specific heat data for pure UBe$_{13}$.

The single crystals of U$_{0.965}$Th$_{0.035}$Be$_{13}$ were grown
from an Al flux as described in Ref. \cite{Smith:92}. 
From zero-field specific heat measurements the upper and lower 
transitions occur at $T_{c1} \! = \! 0.47(5)$~K and 
$T_{c2} \! = \! 0.35(2)$~K, respectively.
The $\mu$SR measurements were carried 
out using a top loading dilution refrigerator on the M15 beam line 
at the TRI-University Meson Facility (TRIUMF), 
Canada and using a $^4$He gas-flow cryostat on the $\pi$M3 beam line 
at the Paul Scherrer Institute (PSI), Switzerland. The crystals were 
mounted on a Ag plate attached to a cold finger. 
The magnetic field ${\bf H}$ was applied parallel to the 
crystallographic $\hat{\bf c}$-axis and transverse to the
initial $\mu^+$-spin polarization direction. As a local spin-$1/2$
probe, the muon is sensitive only to magnetic interactions and 
precesses about the local magnetic field $B_{\mu}$ with a Larmor frequency 
$\omega \! = \! \gamma_{\mu} B_{\mu}$, 
where $\gamma_\mu / 2 \pi \! = \! 13.55342$~MHz/kOe.
The applied field results in a 
uniform polarization of the localized 
U $5f$-moments, which reside at the corners of a cubic lattice.
The Fourier transform of the $\mu^+$-spin precession signal 
in U$_{0.965}$Th$_{0.035}$Be$_{13}$ shows two distinct symmetric lines 
with an amplitude ratio of 1:2. In the time domain, each signal
was best fit by a Gaussian relaxation function
$G(t) \! = \! \exp(-\sigma^2t^2/2)$, where $\sigma$ is the $\mu^+$-spin
depolarization rate. From the amplitude ratio and the frequencies 
of these two signals, we have determined that the $\mu^+$ 
stops at the (0, 0, 1/4) site, half way between nearest-neighbor U atoms.
Muons stopping between U atoms adjoined along the $\hat{\bf c}$-axis 
direction experience a net dipolar field from the $5f$-moments which 
is parallel to ${\bf H}$, and thus precess at a frequency $\omega_{\parallel}$
that is greater than those stopping in Ag (which provide a zero-shift reference frequency). 
On the other hand, twice as 
many $\mu^+$ stop between U atoms adjoined along the $\hat{\bf a}$- and 
$\hat{\bf b}$-axis directions, where the net dipolar field is antiparallel to ${\bf H}$.
These muons precess at a frequency $\omega_{\perp}$ that is lower than those 
stopping in Ag. 

The Knight shift at the two magnetically inequivalent $\mu^+$-sites is given by
\begin{equation}
K_{\parallel, \perp} = \left( 
\omega_{\parallel, \perp} - \omega_{\rm Ag} \right) / \omega_{\rm Ag} \, .
\label{eq:Knight}
\end{equation}
Figure~1 shows measurements of 
the temperature dependence of $K_{\parallel}$ and $K_{\perp}$ below 30~K
at $H \! = \! 10$~kOe and above 2~K at 6~kOe (insets).
The reduction of $K_{\parallel}$ above $T \! \approx \! 50$~K is 
attributed to crystal electric field (CEF) excitations, which have been
inferred from specific heat \cite{Felton:86} and 
NMR spin-lattice relaxation \cite{Clark:88} studies in pure UBe$_{13}$. 
The effect on the hyperfine coupling is observable for both $\mu^+$-sites
from plots of $K$ vs. $\chi_{\rm mol}$ in the normal state (see Fig.~2), 
where $\chi_{\rm mol}$ is the isotropic bulk molar susceptibility.
The plots are essentially linear between 5 and 50~K (where $K$ follows
a Curie-Weiss behavior) and at temperatures above 63~K, with a change of 
slope between the two regions.
The temperature dependence of $\chi_{\rm mol}^{-1}$ is shown in the inset of
Fig.~2 compared with that for proposed CEF spittings of U$^{4+}$ 
$J \! = \! 4$ \cite{Cox:87} and U$^{3+}$ $J \! = \! 9/2$ \cite{Felton:86} manifolds
in cubic symmetry. The CEF models have been corrected by adding
a molecular-field constant of 57~emu/mol, compared 
to 52~emu/mol in CeCu$_2$Si$_2$ \cite{Goremychkin:93}.
Although the data are consistent with $J \! = \! 9/2$,
the $J \! = \!4$ energy scheme results in similar behavior when the hybridization 
proposed in the quadrupolar Kondo model \cite{Cox:87} is included
in the calculation of $\chi_{\rm mol}^{-1}(T)$ --- as was shown for the case of
pure UBe$_{13}$ \cite{McElfresh:93}. A linear fit to $\chi_{\rm mol}^{-1}(T)$ 
above 100~K yields an effective moment of 3.62(1)~$\mu_B$/U. 

Equation~(\ref{eq:Knight}) can be expressed in terms of the individual
contributions to $K$, so that for the axial symmetry of the $\mu^+$-site
\begin{equation}
K_{\parallel}  =  
(A_{\rm c}^{\parallel} + A_{\rm dip}^{zz}) \chi_{5f}
+ K_{\rm dem, L} + K_0 + K_{\rm dia}
\label{eq:KexpPara}
\end{equation}
and
\begin{equation}
K_{\perp}  = 
(A_{\rm c}^{\perp} - \frac{1}{2} A_{\rm dip}^{zz}) \chi_{5f}
+ K_{\rm dem, L} + K_0 + K_{\rm dia} \, \, ,
\label{eq:KexpPerp}
\end{equation}
where $A_{\rm c}$ and $A_{\rm dip}^{zz}$ are the contact hyperfine and dipolar
coupling constants pertaining to the interaction of the $\mu^+$ with
the $5f$-electrons ({\em i.e.} $A \! \equiv \! H_{hf}/N_A \mu_B$, 
where $H_{hf}$ is the hyperfine field, $N_{\rm A}$ is Avogadro's number 
and $\mu_B$ is the Bohr magneton), $\chi_{5f}$ 
is the isotropic molar $5f$-electron susceptibility,
$K_{\rm dem, L} \! = \! 4 \pi(1/3-N) \rho_{\rm mol} \chi_{\rm mol}$ 
is the correction for the demagnetization and Lorentz fields 
(where $N \! \approx \! 1$ is the demagnetization factor and
$\rho_{\rm mol} \! = \! 0.01227$~mol/cm$^3$ is the molar density),
$K_0$ is the isotropic $T$-independent contribution from
the non-$5f$ conduction electrons, and $K_{\rm dia}$ is due to flux
expulsion in the SC state. 

The total normal-state susceptibility is given by
$\chi_{\rm mol} \! = \! \chi_{5f} \! + \! \chi_0$,
where $\chi_0$ is the $T$-independent non-$5f$ 
contribution. From the normal-state plot of 
$K_{\parallel} \! - \! K_{\perp}$ vs. $\chi_{\rm mol}$ at 10~kOe
(see Fig.~3), $\chi_0 \! = \! 0.0039(2)$~emu/mol was obtained from the 
intercept of the extrapolated linear line, where 
$\chi_{5f} \! \propto \! 1/T \! \rightarrow \! 0$
and $K_{\parallel} \! = \! K_{\perp} \! \equiv \! K_0 \! = \! 1846(90)$~ppm.
In general, $A_{\rm c}^{\parallel} \! = \! A_{\rm c}^{\perp}$, in which
case the slope of the solid line (3/2)$A_{\rm dip}^{zz}$ gives 
$A_{\rm dip}^{zz} \! = \! 2066(22)$~Oe/$\mu_B$. 
This value agrees with the result $A_{\rm dip}^{zz} \! = \! 2062$~Oe/$\mu_B$
obtained from a simple dipolar-field calculation for U moments 
sitting on the corners of a cubic lattice of edge-length 5.134~\AA.
We note that the value of $A_{\rm dip}^{zz}$ obtained from the
6~kOe data is only 1693(28)~Oe/$\mu_B$. Although this may imply
that $A_{\rm c}$ is anisotropic, the time spectra recorded at
6~kOe had a larger time resolution and far fewer muon-decay events than
the spectra taken at 10~kOe. Thus, it is likely that there is a systematic 
uncertainty in the temperature dependence of the $\mu^+$-Knight shift at 6~kOe.

$A_{\rm c}$ represents the transferred hyperfine coupling between the 
$\mu^+$-spin and the U $5f$-electrons via the conduction $s$-electrons.
Substituting the value of $A_{\rm dip}^{zz}$ into Eqs.~(\ref{eq:KexpPara}) and 
(\ref{eq:KexpPerp}) gives the temperature dependence of $A_{\rm c}^{\parallel}$ and 
$A_{\rm c}^{\perp}$ shown in Fig.~4. The offset of the 6~kOe data stems from
the discussion in the previous paragraph. The decrease above 50~K is likely 
due to the mixing of the wave functions associated with the different CEF levels.
The strong reduction of $A_{\rm c}^{\perp}$ and lack of change 
of $A_{\rm c}^{\parallel}$ for $T_{c1} \lesssim \! T \! \lesssim \! 4$~K 
is the source of the nonlinearity above 
$\chi_{\rm mol} \! \approx \! 0.014$~emu/mol in Fig.~3.
This is not a muon induced effect, since
similar departures from linearity have been observed in
$K$-$\chi_{\rm mol}$ plots for the $^9$Be NMR Knight shift 
in UBe$_{13}$ \cite{Clark:87} and the $^{63}$Cu and $^{29}$Si NMR Knight shifts 
in CeCu$_2$Si$_2$ \cite{Ohama:95}. A decrease 
of $A_{\rm c}^{\perp}$ over nearly the same temperature range is
also observed in pure UBe$_{13}$ \cite{Sonier:00}.
This anomaly roughly coincides with the peak in the resistivity 
and specific heat at $\sim \! 2.5$~K in UBe$_{13}$ \cite{Mayer:86}, 
which is completely suppressed when 3.55~\% Th is added.
The decrease of $A_{\rm c}^{\perp}$ in both the pure and doped
systems is not necessarily inconsistent with this latter behavior,
because most of the $\mu^+$ stopping in U$_{0.965}$Th$_{0.035}$Be$_{13}$
do not reside near a Th impurity.

In the SC state the flux expulsion 
term $K_{\rm dia}$ in Eqs.~(\ref{eq:KexpPara}) 
and (\ref{eq:KexpPerp}) is related to the value of the magnetic
penetration depth $\lambda$ and the coherence length $\xi_0$.
To our knowledge there have been no measurements of the
absolute value of $\lambda$ in U$_{0.965}$Th$_{0.035}$Be$_{13}$.
However, the lack of any increase in the $\mu^+$-spin depolarization
rate $\sigma$ below $T_{c1}$ is consistent with a value
$\lambda(0) \! \gg \! 12100$~\AA, as reported 
in pure UBe$_{13}$ \cite{Reotier:00}. Using the simple theoretical 
model developed by Hao {\em et al.} for the reversible 
magnetization of a type-II superconductor \cite{Hao:91} and
the value $H_{c2}(0) \! \approx \! 55$~kOe \cite{Schmied:88},
we calculate that $| K_{\rm dia} | \! \ll \! 72$~ppm.
Since we observed no field dependence for $K_{\parallel,\perp}$ below
$T_{c1}$ in the range 5~kOe~$\! \leq \! H \leq \! 15$~kOe, 
we conclude that the internal field is essentially uniform and the diamagnetic 
shift $K_{\rm dia}$ is negligible.

Because $A_{\rm c}^{\parallel}$ is temperature independent in the
normal state, we make the reasonable assumption that it
remains so below $T_{c1}$, allowing $\chi_s$ ({\em i.e.}, 
$\chi_{5f}$ in the SC state) to be determined 
from Eq.~(\ref{eq:KexpPara}). 
As shown in Fig.~5, $\chi_s(T)$ exhibits two different behaviors 
(in agreement with the raw Knight shift data in Fig.~1) which 
coincide with the two phase transitions in the specific heat.
The decrease of $\chi_s(T)$ between $T_{c1}$ and $T_{c2}$ 
is consistent with a phase in which the Cooper pairs 
have a substate of opposite spin projection 
({\em i.e.}, $S_{z} \! = \! 0$). However, the data cannot 
distinguish between even and odd parity spin states 
possessing this substate, because Fermi-liquid 
corrections and spin-orbit (SO) scattering by impurities may be
significant. For the case of an even parity SC phase we can estimate the 
importance of SO scattering from the relation 
$\chi_s(T_{c2})/\chi_n \! = \! 1 \! - \! 2 l_{\rm SO}/\pi \xi_0$ 
\cite{Anderson:59}, where $\chi_n$ is the normal-state 
spin susceptibility at $T_{c1}$, $\chi_s(T_{c2})/\chi_n \! = \! 0.61$
and $\xi_0 \! \approx \! 77$~\AA~ from $H_{c2}(0)$ \cite{Schmied:88}.
This gives a SO scattering mean free path of $l_{\rm SO} \! \approx 47$~\AA. 
The average distance between Th atoms $\approx \! 15$~\AA, 
represents a lower limit for the mean free path $l$ between collisions of the 
electrons with the Th impurities. Since $l_{\rm SO}$ is of the same order of 
$l$, modification of $\chi_s(T)$ due to SO scattering cannot be ruled out.

The lack of a temperature dependence for $\chi_s$ below 
$T_{c2}$ is characteristic of a spin-triplet ($S \! = \! 1$)
odd-parity ($L \! = \! 1$) superconductor with 
{\em parallel spin pairing}, except that 
$\chi_s \! < \! \chi_n$. 
This unusual behavior suggests that the component of the
order parameter corresponding to the phase
$T_{c2} \! < \! T \! < \! T_{c1}$ stops or slows down its
growth at $T_{c2}$, where a second component develops.
In terms of the ${\bf d}$-vector \cite{Vollhardt:90} of the 
triplet order parameter 
$\hat{\Delta}({\bf k}) \! = \! i({\bf d} \! \cdot \! {\bf \sigma})\sigma_y$, 
a possibile scenario is that
(i) one component corresponds to ${\bf d} \! \parallel \! {\bf H}$,
so that $\chi_s$ decreases below $T_{c1}$ and (ii) the second component
corresponds to ${\bf d} \! \perp \! {\bf H}$, in which case $\chi_s$
is unchanged below $T_{c2}$. The idea of a two-component
${\bf d}$-vector is similar to the weak spin-orbit coupling model
recently developed for UPt$_3$ \cite{Machida:98} 
from detailed $^{195}$Pt NMR Knight 
shift measurements \cite{Tou:96}. 
Finally, substituting $\chi_s(T)$ for $\chi_{5f}$ in Eq.~(\ref{eq:KexpPerp}) 
we find that the magnitude of $A_{\rm c}^{\perp}$ rapidly increases to a constant
value below $T_{c2}$ (see Fig~4).

In conclusion, our study of U$_{0.965}$Th$_{0.035}$Be$_{13}$ 
has identified different behavior for the temperature 
dependence of the spin susceptibility $\chi_s$ below the 
two transitions observed in the specific heat. 
A possible explanation for the absence of a change below $T_{c2}$ is that 
U$_{0.965}$Th$_{0.035}$Be$_{13}$ is an odd parity spin-triplet 
superconductor. However, we stress that this may not be the
only interpretation of our measurements.
A definitive identification of the pairing state will 
require further measurements as a function of 
magnetic-field direction to unambiguously determine the 
relative orientation of the ${\bf d}$-vector and ${\bf H}$.
     
We are extremely grateful to K.~Machida, D.L.~Cox, 
M.J.~Graf, A.V.~Balatsky and A.~Schenck for informative discussions.
Work at Los Alamos was performed under the auspices of the U.S. DOE.
Other support was provided by U.S. NSF, Grant DMR-9731361 
(U.C. Riverside), DMR-9820631 (California State) and the Dutch 
Foundations FOM and NWO (Leiden).    

%%%%%%%%%%%%%%%%%%%%%%%%%%%%%%%%%%%%%%%%%%%%%%%%%%%%%%%%%%%%%%%%%%%%%%%%%%%% 

%%% References should come before figure captions %%%

% REFERENCES 

%%%%%%%%%%%%%%%%%%%%%%%%%%%%%%%%%%%%%%%%%%%%%%%%%%%%%%%%%%%%%%%%%%%%%%%%%%% 
\newpage 

\begin{center} 
FIGURE CAPTIONS 
\end{center} 
 
Figure 1. Temperature dependence of 
$K_{\parallel}$ and $K_{\perp}$ measured at TRIUMF in an applied 
field $H \! = \! 10$~kOe. Inset: Data taken above $T_{c1}$ at PSI, 
where the maximum available field was $H \! = \! 6$~kOe.\\

Figure 2. Plot of the normal-state $\mu^+$-Knight shift
at $H \! = \! 6$~kOe vs. the bulk molar susceptibility.  
Inset: Temperature dependence of the inverse susceptibility. 
Dashed and solid lines are calculations 
using Eq.~(3) of Ref.~\cite{Pagliuso:99}
for the CEF schemes described in the text.\\

Figure 3. Plot of $K_{\parallel} \! - \! K_{\perp}$
vs. the bulk molar susceptibility at $T \! > \! T_{c1}$
and $H \! = \! 10$~kOe. 
The solid line is a linear fit to the data above 5~K.\\

Figure 4. Temperature dependence of $A_{\rm c}^{\parallel}$ (open symbols) and
$A_{\rm c}^{\perp}$ (solid symbols) at $H \! = \! 6$~kOe and
10~kOe. Note: We have assumed that $A_{\rm c}^{\parallel}$ is unchanged 
below $T_{c1}$. The data for $A_{\rm c}^{\perp}$ below $T_{c1}$ was 
obtained under this assumption.\\ 

Figure 5. Temperature dependence of the specific heat (open circles)
and magnetic susceptibility (solid circles). 

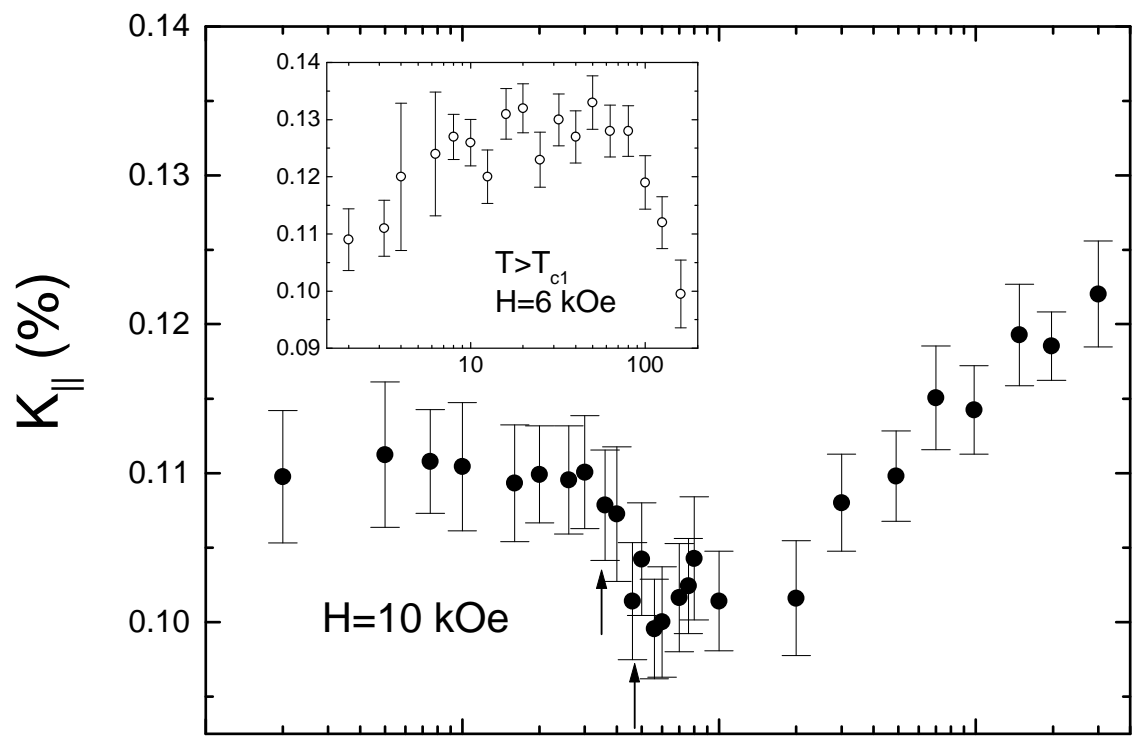
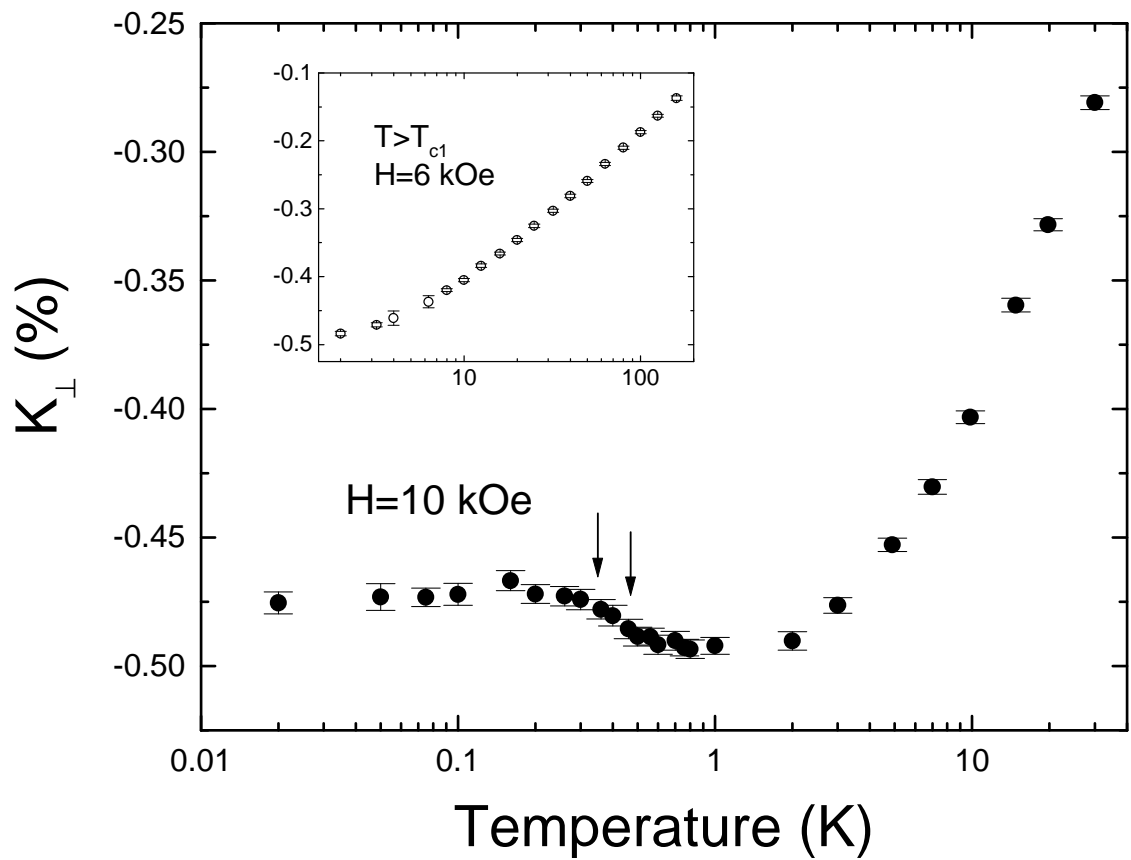
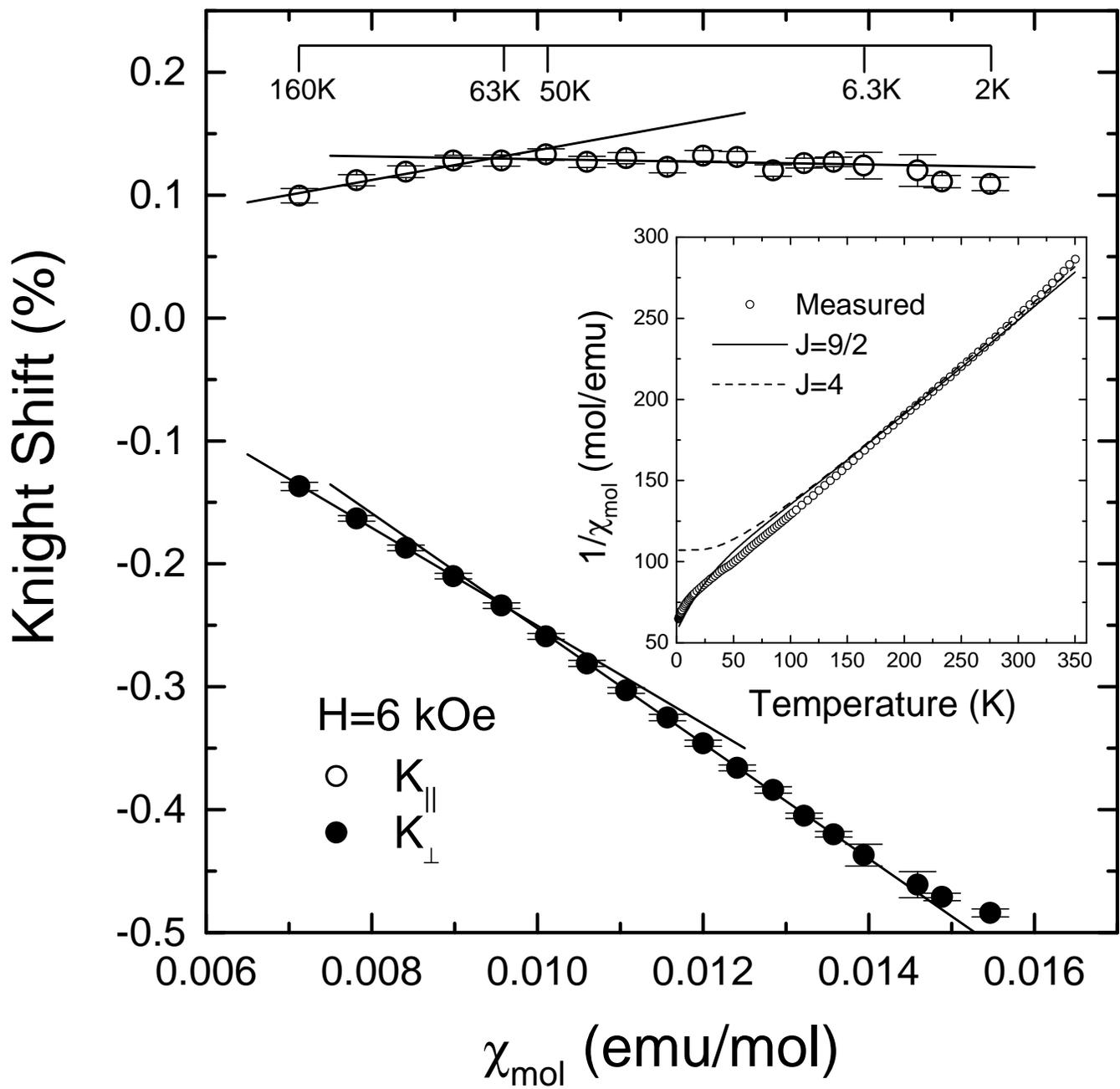
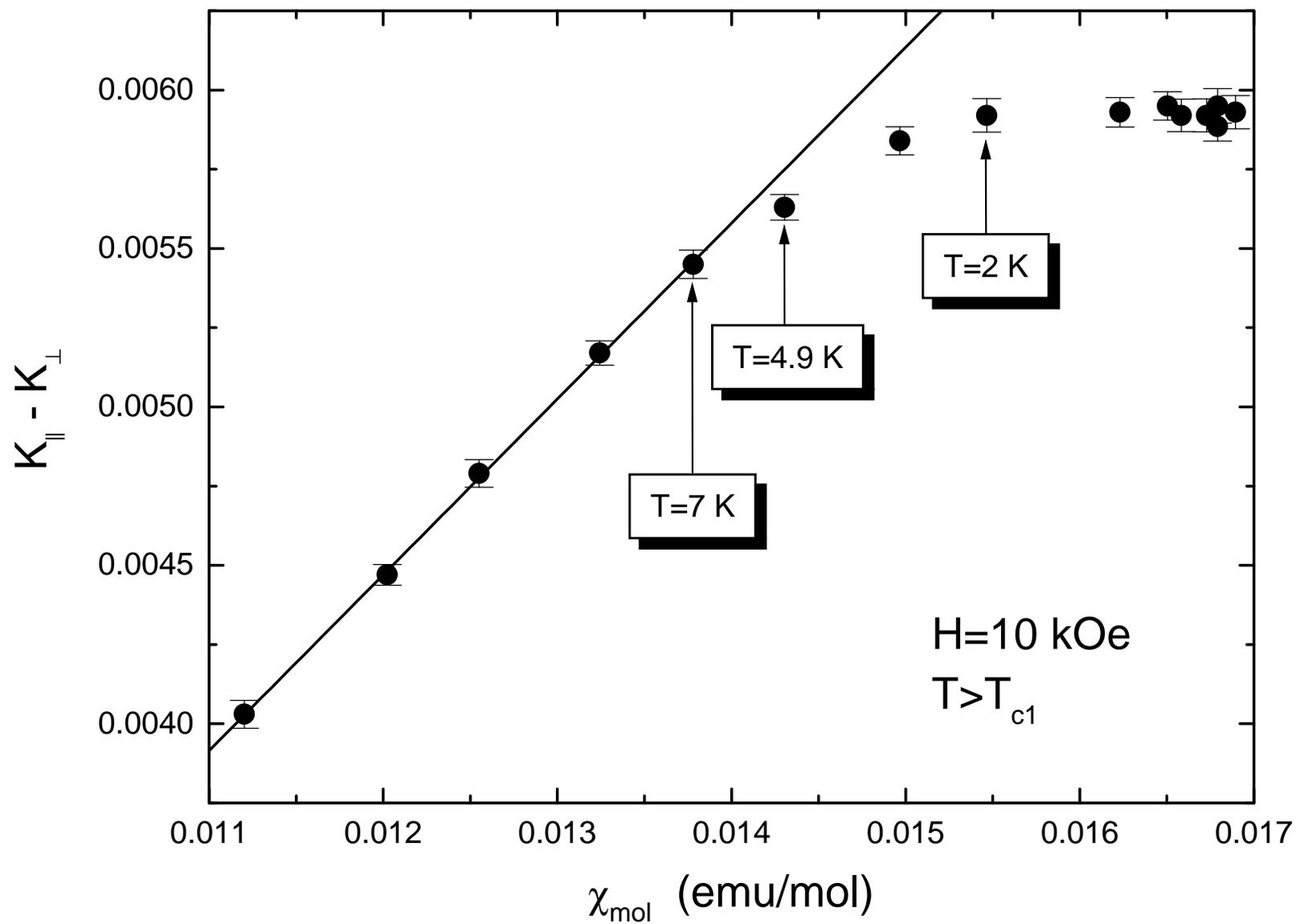
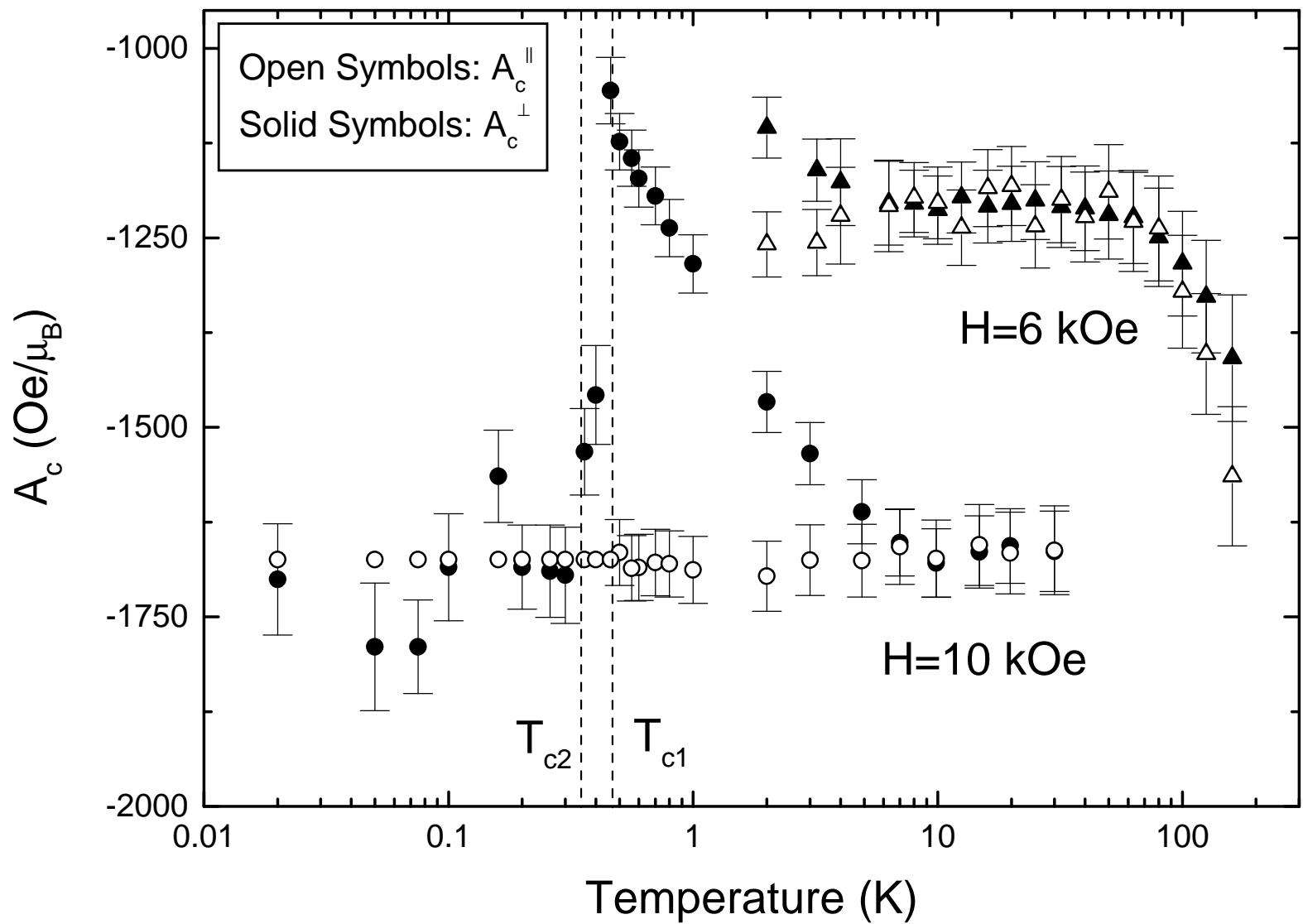
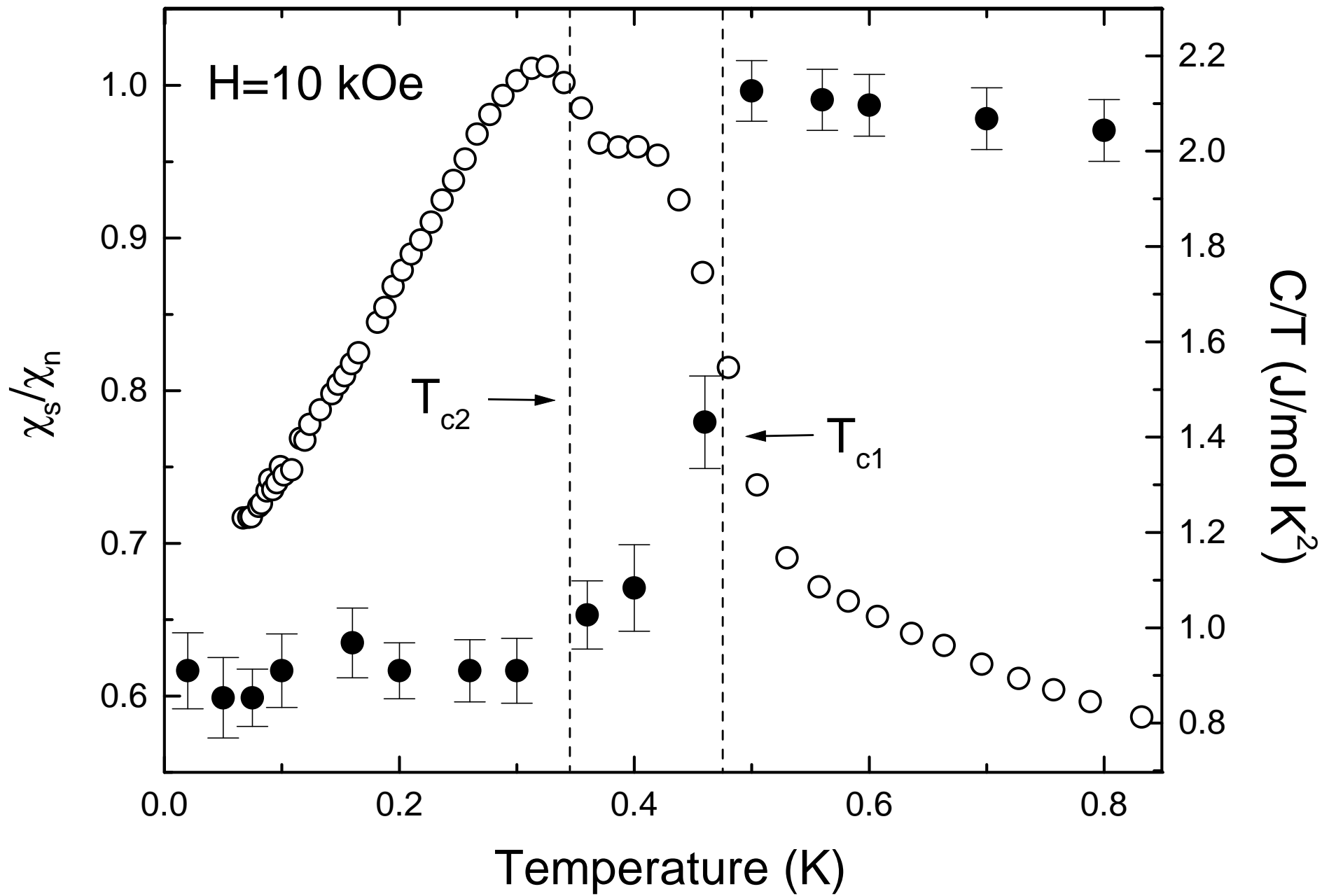
\end{document}